\documentclass[review]{elsarticle}
\usepackage{graphicx} 
\usepackage{amsmath}
\usepackage{graphicx}
\usepackage{tabularx}
\usepackage{longtable}
\usepackage{geometry}
\usepackage{algorithm}
\usepackage{algpseudocode}
\usepackage{lineno,hyperref}
 \usepackage{amsmath} 
 \usepackage{multirow} 
 \usepackage{booktabs} 
 \usepackage{qcircuit}

\allowdisplaybreaks[3]

\journal{Elsevier}

\makeatletter
\newenvironment{breakablealgorithm}
  {
   \begin{center}
     \refstepcounter{algorithm}
     \hrule height.8pt depth0pt \kern2pt
     \renewcommand{\caption}[2][\relax]{
       {\raggedright\textbf{\ALG@name~\thealgorithm} ##2\par}%
       \ifx\relax##1\relax 
         \addcontentsline{loa}{algorithm}{\protect\numberline{\thealgorithm}##2}%
       \else 
         \addcontentsline{loa}{algorithm}{\protect\numberline{\thealgorithm}##1}%
       \fi
       \kern2pt\hrule\kern2pt
     }
  }{
     \kern2pt\hrule\relax
   \end{center}
  }
\makeatother

\begin{document}

\begin{frontmatter}

\title{Exact Quantum Algorithm for Unit Commitment Optimization based on
Partially Connected Quantum Neural Networks}


\author[mysecondaddress]{Jian Liu}
\ead{shjtdxlj@hotmail.com}

\author[mymainaddress,mysixthaddress]{Xu Zhou\corref{mycorrespondingauthor}}
\ead{zhoux359@mail.sysu.edu.cn}

\author[mysecondaddress]{Zhuojun Zhou}
\ead{zjzhou@qudoor.cn}

\author[mymainaddress,1,3]{Le Luo}
\ead{luole5@mail.sysu.edu.cn}

\cortext[mycorrespondingauthor]{Corresponding author}

\address[mymainaddress]{School of Physics and Astronomy, Sun Yat-sen University, Zhuhai 519082, China}

\address[mysecondaddress]{QUDOOR Co, Ltd., Hefei 230000, China}

\address[mysixthaddress]{QUDOOR Co, Ltd., Beijing 100089, China}

\address[1]{Shenzhen Research Institute of Sun Yat-Sen University, Shenzhen 518057, China}

\address[3]{Guangdong Provincial Key Laboratory of Quantum Metrology and Sensing, Sun Yat-Sen University, Zhuhai 519082}

\begin{abstract}
The quantum hybrid algorithm has become a 
very promising and speedily method today for solving the larger-scale optimization in the noisy intermediate-scale quantum (NISQ) era. The unit commitment (UC) problem is a fundamental problem in the power system which
aims to satisfy a balance load with minimal cost.
In this paper, we focus on the implement of the UC-solving by exact quantum algorithms based on the quantum neural network (QNN).
This method is tested with up to 10-unit system with the balance load constraint. 
In order to improve the computing precision and reduce the network complexity, we suggest the knowledge-based partially connected quantum neural network (PCQNN).
The results show that the exact solutions can be obtained by the improved algorithm and the depth of the quantum circuit can be reduced simultaneously.
\end{abstract}


\end{frontmatter}


\section{Introduction}
Unit commitment (UC) is a critically important optimization problem in the electrical power system, which aims to meet the user's demand for electric energy by rationally scheduling the generator sets in the power system.
The UC problem is to determine the most appropriate power unit start-up and shut-down schedule to minimize the cost of power generation while meeting a large number of operating and security constraints.
It is a nonlinear mixed-integer quadratic program (MIQP) with both continuous and integer variables. The treatment of the binary commitment decision variables induces large combinatorial complexity.
Numerous methods, such as those referenced in \cite{cheng2000unit} and \cite{haoyong1999evolutionary}, have been employed to tackle this NP-hard problem. While algorithms like branch-and-bound theoretically offer optimal solutions to the UC problem, their application in large-scale systems is limited due to extensive computation times. Other approaches, including priority lists \cite{moussouni2007optimization} and Lagrangian relaxation \cite{burns1975optimization}, do not guarantee optimal solutions. Hybrid algorithms, such as genetic algorithms \cite{zhuang1988towards} and machine learning \cite{kazarlis1996genetic}, have been proposed to combine different methods and enhance solution quality.

Although some quantum algorithms have been greatly developed for optimization problems in recent years, the number of qubits in quantum computers is still limited. So how to solve larger-scale problems in the noisy intermediate-scale quantum (NISQ) era has become one of the most urgent problems. The design, implementation and application of new quantum algorithms toward lager system have become the focus of attention.
The variational quantum algorithm (VQA) \cite{cerezo2021variational} is one of the most promising schemes in the NISQ algorithms, including variational quantum eigensolver (VQE) \cite{li2019variational}, quantum approximate optimization algorithm (QAOA) \cite{farhi2016quantum}, quantum neural network (QNN) \cite{li2022quantum,kwak2021quantum} et al.
VQA has been developed for applications with a wide range, including calculating molecular ground states, dynamical simulation of many-body systems and solving linear equations.
Recently, VQE and QAOA \cite{koretsky2021adapting,mahroo2022hybrid,stein2023combining,nikmehr2022quantum} were both proposed to tackle the UC problem. To deal with continuous variables, some classical aided methods were used with the quantum algorithm together, such as alternating direction method of multipliers (ADMM) \cite{yang2023quantum} and the Benders decomposition \cite{benders} et al.

In our research, we propose the knowledge-based Partially Connected Quantum Neural Network (PCQNN) algorithm to achieve optimal efficiency in solving the UC problem.
Fully connected neural networks (FCNNs) are prevalent in classical neural network applications. However, they often include a significant amount of redundancy. In contrast, a Partially Connected Neural Network (PCNN) is characterized by a network that includes only a subset of the potential connections present in a complete neural network model, as surveyed in \cite{elizondo1997survey}.
The concept of PCQNN is to simplify the topology of a quantum neural network based on symbolic domain knowledge. This approach aims to enhance the performance of a hardware-efficient ansatz (HEA) model while utilizing a reduced number of connections, thereby increasing computational efficiency.

In Section \ref{sec2}, we initially construct the theoretical framework for the optimization problem, including both equality and inequality constraints. Subsequently, the UC problem is translated into an interactional Ising Hamiltonian model.
Moving to Section \ref{sec3}, we detail the QNN circuit designed for a 10-unit system with a single constraint optimization. Following this, we introduce a problem-inspired PCQNN circuit for a 5-unit system with double constraint optimization. Our findings presented in Section \ref{sec4} demonstrate that, similar to PCNNs in classical computing, the PCQNN circuit can achieve higher precision with a shallower circuit depth in quantum optimization tasks.
Consequently, our research proposes an accurate algorithm that is well-suited for the UC problem, particularly in the distributed quantum computing environments. Moreover, this methodology is not confined to UC problems. It possesses the potential to be extended to a broad spectrum of combinatorial optimization problems.

\section{Theoretical framework}\label{sec2}

A general optimization problem can be written mathematically as follows,
\begin{eqnarray}
    \text{minimize} \quad  f_0(x), \\
    \text{subject to} \quad  f_{i,eq}(x)=A_i, \\
     f_{i,neq}(x) \le B_i,
\end{eqnarray}
where $f_0:\textbf{R}^n\to \textbf{R}$ is the objective function, $f_{i,eq},f_{i,neq}:\textbf{R}^n\to \textbf{R}$ are equality and inequality constraints respectively, and $x=(x_1,x_2,\dots,x_n )$ are optimization binary  variables with domains of $x_i \in \left \{ 0,1 \right \} $, $A_i$ and $B_i$ are constants. Main parameters used in this paper are shown within Table~\ref{para}.

\begin{table}[t]
    \caption{Parameters used in this paper}
    \label{param}
    \begin{center}
    \begin{tabular}{c|c|c|c}
    \toprule
        \textbf{Symbol} & \textbf{Description} & \textbf{Symbol} &  \textbf{Description}   \\
\hline
         $f_0$ & objective function & $a_i$ & quadratic coefficients  \\
         $f_{i,eq}$ & equality constraints & $b_i$ & linear coefficients \\
         $f_{i,neq}$ & inequality constraints & $c_i$ & opening costs \\
         $x_i$ & binary variables & $p_i$ & power of $i$th unit \\
         $H$ & Hamiltonian & $L$ & total load  \\
         $Z_i$ & Pauli Z operators & $S$ & system spare power \\
         $s_i$ & relaxation variables & $\theta$ & variational parameters \\
         $n$ & number of unit qubits & $\theta^*$ & optimization parameters \\
         $k$ & number of auxiliary qubits & $d$ & number of the layers \\
         $\lambda_i$ & weight of penalty functions & & \\
\bottomrule
    \end{tabular}
    \end{center}
    \label{para}
\end{table}

The optimization problem should be transformed into the Quadratic Unconstrained Binary Optimization (QUBO) problem with the Hamiltonian of Ising model so that it can be solved in quantum computers,
\begin{eqnarray}
     H =\sum_{i=1}^{n}\mathcal{C}_{ii}  Z_i +\sum_{1\le i<j\le n}^{n}\mathcal{C}_{ij} Z_iZ_j,\label{Ham1}
\end{eqnarray}
where $\mathcal{C}_{ii}$ and $\mathcal{C}_{ij}$ are undetermined coefficients, $Z_i$ are Pauli operators for spin qubits. This process can be achieved by following steps.

\begin{itemize}
    \item Transform inequality constraints into equality constraints by adding relaxation variables $s_i$,
    \begin{eqnarray}
        f_{i,neq}(x)-B_i-s_i=0, \quad i=1,\dots ,m,
    \end{eqnarray}
    where integer variables $s_i$ can be represented efficiently as a binary problem, requires a logarithmic number of binary variables,
    \begin{eqnarray}
        s_i=\sum_{j=1}^{k}2^{j-1}x_{n+j}.\label{binary}
    \end{eqnarray}
A integer variable $s_i$ from 0 to N, requires binary variables $x_j$ of the number,
\begin{eqnarray}
    k=\text{log}_2N.
\end{eqnarray}

      \item Obtain a quadratic unconstrained binary optimization (QUBO) problem by penalty function methods,
    \begin{eqnarray}
        f_0(x)+\lambda _1(f_{i,eq}(x)-A_i)^2+\lambda _2(f_{i,neq}(x)-B_i-s_i)^2, \label{QUBO}
    \end{eqnarray}
    where $\lambda_i$ is the weight of penalty function.

    \item Transform binary variables into spin by
        \begin{eqnarray}
       x_i=\frac{I-Z_i}{2}, \quad i=1,\dots ,n+k, \label{binary-spin}
    \end{eqnarray}
    where $Z_i$ are the Pauli $Z$ operations. By submitting Eq.\eqref{binary-spin} into 
     Eq.\eqref{QUBO}, the Hamiltonian with the form of Eq.\eqref{Ham1} can be obtained by calculation.    
\end{itemize}

From Eq.\eqref{binary} we can find that, in order to deal with the relaxation variables $s_i$, one needs $k$ auxiliary qubits. Thus inequality constraints may induce additional complexity within the computing. In oder to reduce the number of the auxiliary qubits, it is necessary to perform a case by case work.

Specifically, the UC solution determines
the combination of units’ on/off status to meet the load at a
time period. For a single time period, the objective function shall be the sum of the cost of the opened generator units,
\begin{eqnarray}
  \sum_{i=1}^{n}f_i x_i= \sum_{i=1}^{n} (a_ip_i^2+b_ip_i+c_i) x_i,
\end{eqnarray}
where $a_i$, $b_i$ and $c_i$ are the fixed cost coefficients of $i$th unit.

For an $n$-unit power system, the system power balance shall be preserved  by an equality constraint,
\begin{eqnarray}
    \sum_{i=1}^{n}p_ix_i=L ,
\end{eqnarray}
 where $L$ is the total load.
 The system spare constrain can be expressed by the following inequality,
 \begin{eqnarray}
    \sum_{i=1}^{n}p_{i,max}x_i\ge L +S,
\end{eqnarray}
where $S$ is the system spare power. 
By transforming the constrains into penalty functions and using the binary-spin transformation, we can obtain the Hamiltonian for the specific problem,

\begin{eqnarray}
    H&=&\sum_{i=1}^{n}f_{i}\frac{I-Z_i}{2} +\lambda _1\left ( \sum_{i=1}^{n}p_{i}\frac{I-Z_i}{2} -L  \right ) ^2+ \nonumber \\
    &&\lambda _2\left ( \sum_{i=1}^{n}p_{i,max}\frac{I-Z_i}{2} -L -S-c_c\sum_{j=1}^{k}2^{j-1}\frac{I-Z_{n+j}}{2}   \right ) ^2, \label{Ham2}
\end{eqnarray}
where $c_c$ is a constant determined by the magnitude of relaxation variables. 

\section{Quantum algorithm for UC} \label{sec3}
In this section, we firstly describe how a basic quantum neural network works, and then induce the PCQNN circuit for the UC problem.

\subsection{Quantum Neural Networks}
The way of hardware-efficient ansatz
QNN processes data is as follows.

\begin{itemize}
\item Encode the input data into the corresponding quantum states with an appropriate number of qubits.
\item The quantum state of all qubits is transformed by passing parameterized rotation gates and entanglement gates with repeating a given number of layers.
\item Measure the output quantum state and obtain the expectation value of Hamiltonian operators which is consisted of Pauli gates. 
\item The parameters are updated by an optimizer, such as ADAM, BFGS et al, based on the feedback loop of the QNN circuit.
\end{itemize}

\begin{figure}[H]
\centering
\includegraphics[width=0.7\textwidth]{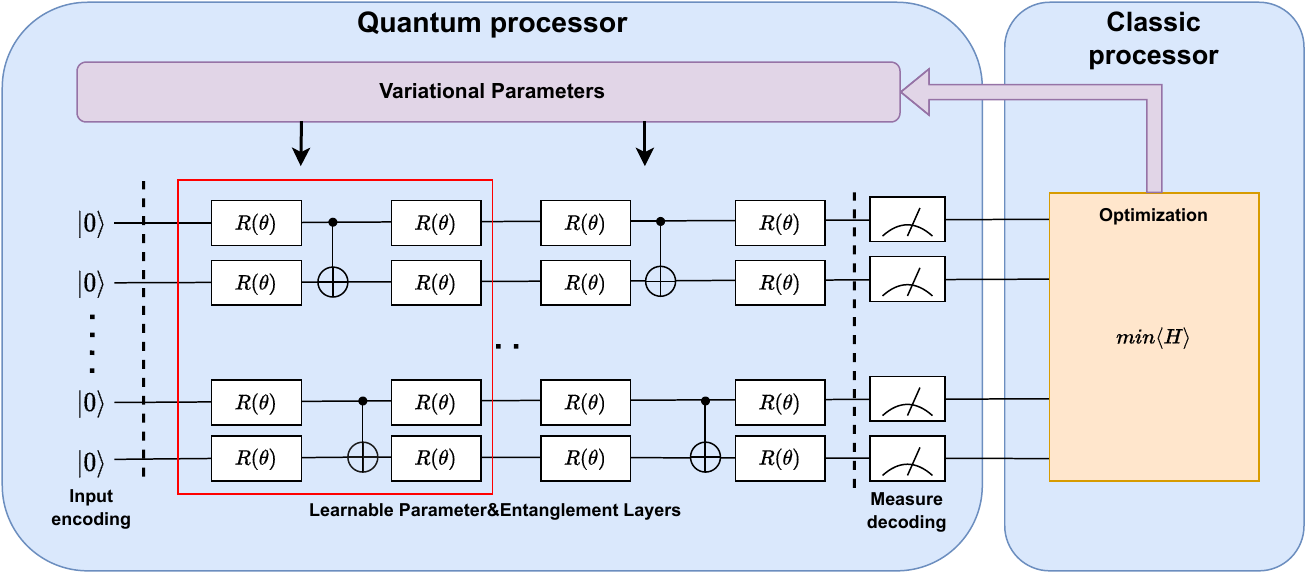}
\caption{Diagram of the QNN quantum-classical algorithm. The variational parameters are $\theta=(\theta_1,\theta_2, \cdots)$. The rotation gates $R$ can be decomposed to the rotations in 3D, $R_x$, $R_y$, $R_z$. The red frame denotes one repeated layer.}
\label{fig1}
\end{figure}

Fig.\ref{fig1} shows the example of a QNN algorithm.

\subsection{Knowledge-based PCQNN circuit}
In order to improve the efficiency of the circuit operation, we apply the PCQNN by a knowledge-based method. The idea is to use symbolic domain knowledge as rules to actually create the topology of the neural
network. Next, the actual neural training quantum algorithm is applied to improve the performance of the network.

We first induce some parameters for the UC problem, including the total cost $\mathcal{F}$, total power $\mathcal{P}$, total max power $\mathcal{M}$ and the relaxation variable $\mathcal{R}$, which can be defined as
\begin{eqnarray}
    \mathcal{F}&=&\sum_{i=1}^{n} f_ix_i, \\ \mathcal{P}&=&\sum_{i=1}^{n} p_ix_i, \\
    \mathcal{M}&=&\sum_{i=1}^{n} p_{max}x_i, \\
    \mathcal{R}&=&\sum_{j=1}^{k}2^{j-1}x_{n+j}.
\end{eqnarray}

Therefore, the Hamiltonian of the UC problem 
can be rewrote as
\begin{eqnarray}
   H=\mathcal{F}  +\lambda _1(\mathcal{P}-L)^2+\lambda _2(\mathcal{M}-\mathcal{R}-L-S)^2.
\end{eqnarray}

We have identified that the initial three parameters are indicative of the physical characteristics of the generator units. These parameters can be determined through the arguments associated with the initial $n$ qubits. In contrast, the last parameter represents the relaxation variable, which can be determined by the subsequent $k$ qubits. Consequently, the UC problem can be approached as an optimization involving the symbolic parameters $\mathcal{F}$ (fuel cost), $\mathcal{P}$ (power output), $\mathcal{M}$ (maintenance), and $\mathcal{R}$ (relaxation).

\begin{figure}[t]
\centering
\includegraphics[width=0.6\textwidth]{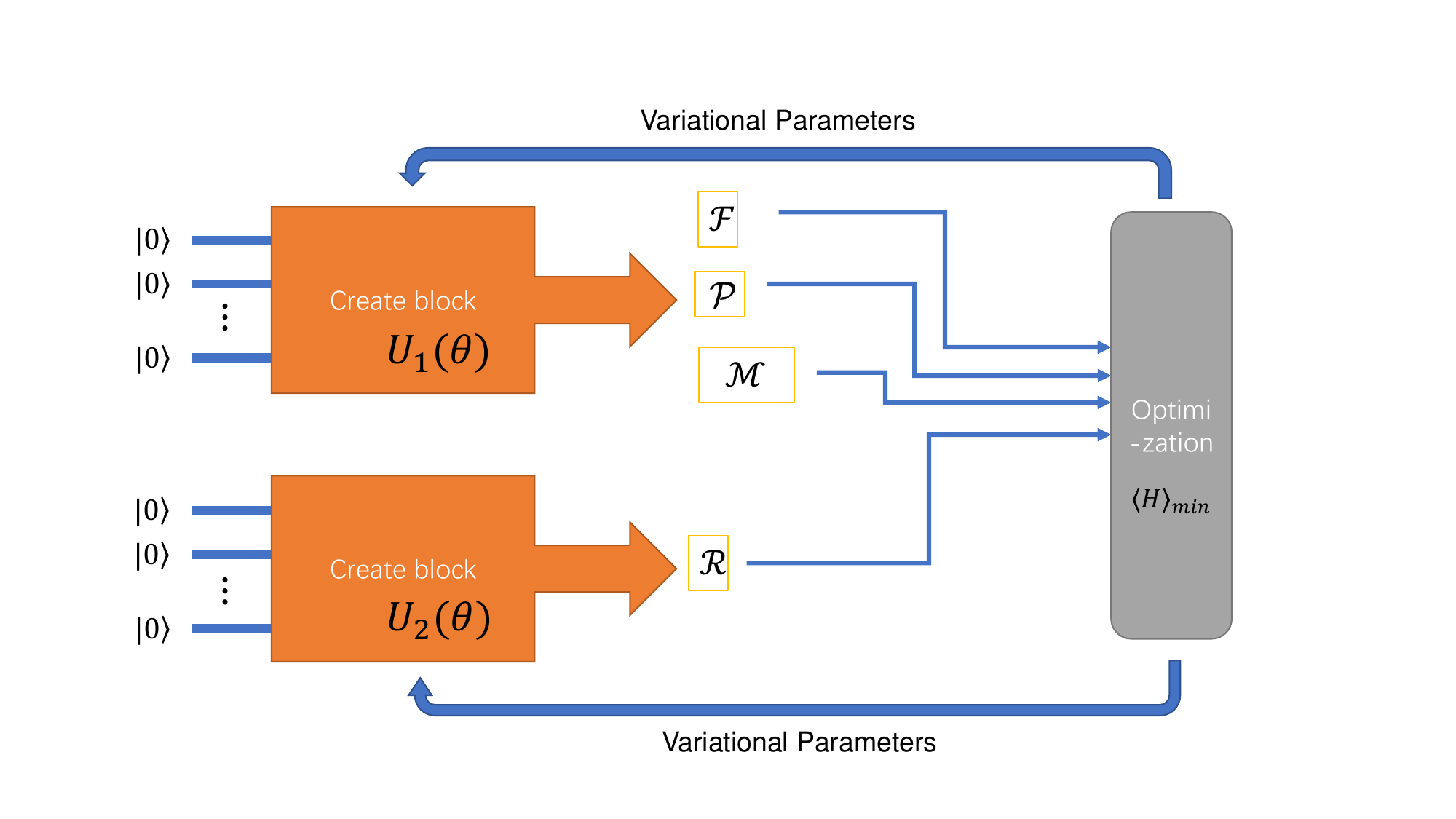}
\caption{Diagram of the quantum algorithm of the
knowledge-based
PCQNN for the UC problem.}
\label{PCQNNfig}
\end{figure}

Consequently, one can input the generator's properties and the relaxation variable into the network as two distinct sets of data. Subsequently, the information from these two inputs can be integrated at the end of the quantum network's processing by a classical optimizer. This approach ensures a balanced focus on both the local details and the overarching system, thereby achieving the desired optimization goals. A simple schematic diagram is shown in Fig.\ref{PCQNNfig}.

Our PCQNN circuit with the problem-inspired ansatz can be obtained by directly removing entangled gates from the original HEA QNN circuit (as shown in Fig.\ref{fig2}), since they can be generated independently from each other without any entanglement between them. The procedure can be found in the Algorithm \ref{alg:PCQNN}.

\begin{breakablealgorithm}
\caption{knowledge-based PCQNN algorithm}
\label{knowledge-based PCQNN algorithm}
\begin{algorithmic}
\noindent \textbf{Input}: \\
(1) $n$ : Number of unit nodes\\
(2) $k$ : Number of auxiliary qubits\\
(3) $p_i$ : List of power of $i$th units\\
(4) $p_{i,max}$ : List of max power of $i$th units\\
(5) $d$ : Number of layers in PCQNN\\
(6) $\theta_i$ : List of arguments that will be necessary to build the cost layer 

\noindent \textbf{Output}: The target string $x=(x_1,x_2,\dots,x_{n+k} )$ with great probability

\noindent\textbf{Procedure:}

\noindent \textbf{Step 1.} Create list of $\theta_i$ parameters 

\noindent \textbf{Step 2.} Create list of $n+k$ parameterized rotation gates in 3 dimensions $R_y$, $R_z$ and $R_x$

\noindent \textbf{Step 3.}  Apply all-zero state on all the qubits to initialize the qc (QuantumCircuit)

\noindent \textbf{Step 4.} 
\State $j \leftarrow 1$
    \While{$j \leq d$}
    \State Build and apply cost layer $U_1(\theta)$ for the qubits from 1 to n
    \State Build and apply cost layer $U_2(\theta)$ for the qubits from n+1 to n+k
\EndWhile\label{QNNwhile}

\noindent \textbf{Step 5.}  Measure all qubits in standard basis and store results in classical registers

\noindent \textbf{Step 6.} Use a classical optimizer to find parameters $\theta^*$, to minimize expected cost

\noindent \textbf{End procedure}
\end{algorithmic}
\label{alg:PCQNN}
\end{breakablealgorithm}

\section{Simulation results} \label{sec4}
The $d$ layers of parametrized quantum gates are then
ready to go through the closed loop optimization process with a
classical optimizer. The classical optimizer is used to obtain
variational parameters $\theta^*$ such that the expected function $\left \langle H \right \rangle $ is minimized:
\begin{eqnarray}
    \theta ^*=argmin \mathcal{H} (\theta ),
\end{eqnarray}
where $\mathcal{H}$ is the expectation value,
\begin{eqnarray}
    \mathcal{H(\theta )} =\left \langle\phi (\theta ) |H|\phi(\theta ) \right \rangle,
\end{eqnarray}
with $\phi(\theta)$ the output quantum state after passing the ansatz block of the quantum circuit.

\subsection{10-unit exact solution}
\begin{figure}[H]
\centering
\includegraphics[width=0.7\textwidth]{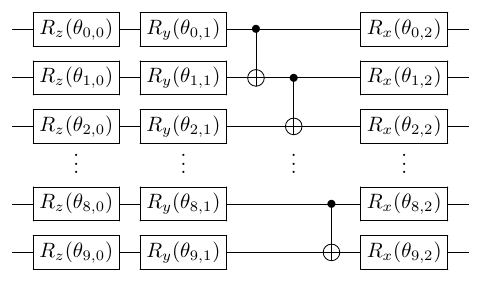}
\caption{Diagram of the quantum circuit of the HEA QNN for 10-unit system and one constraint (one layer).}
\label{fig2}
\end{figure}

The method is tested with up to 10 power units with the power balance constraint firstly. The generator's data can be found in the Table~\ref{10data}. According to the discussion above, the single layer of the QNN circuit is constructed as shown in Fig.\ref{fig2}.
We tested several quantum circuits with different depths and
found that when $d=10$, the solution obtained by quantum computation is slightly different from classical solver, which can be found in the Table~\ref{T1} \footnote{Data of generators, reference: Samantha Koretsky et al, Adapting Quantum Approximation Optimization Algorithm (QAOA) for Unit Commitment, 2021 IEEE International Conference on Quantum Computing and Engineering (QCE)
}.

\begin{table}[H]
    \caption{Data for 10-unit power system.} 
    \begin{center}
    \scalebox{0.85}{
    \begin{tabular}{c|c|c|c|c|c|c|c|c|c|c}
\toprule
     $i$ & 1 & 2 & 3 & 4 & 5 & 6 & 7 & 8 & 9 & 10 \\
\hline
        $a_i(\$/MW^2)$& 0.0048& 0.00031 &0.002 &0.00211&0.00398 & 0.00712 & 0.0079 & 0.00413 & 0.00222 &0.00173  \\
       $b_i(\$/MW)$ & 16.19 & 17.26 &  16.60 & 16.50 & 19.70 & 22.26 & 27.74 & 25.92 & 27.27 & 27.79 \\
        $c_i(\$)$ & 1000 & 970 & 700 & 680 & 450 & 370 & 480 & 660 & 665 & 670  \\
         $p_i(MW)$ & 150 & 150 & 20 & 20 & 25 & 20 & 25 & 10 & 10 & 10   \\
\bottomrule
    \end{tabular}}
    \end{center}
    \label{10data}
\end{table}

\begin{table}[H]
    \caption{Data for 5-unit power system.}
    \begin{center}
    \begin{tabular}{c|c|c|c}
    \toprule
    \multirow{2}{*} {$L(MW)$} & \multirow{2}{*} {\textbf{Classical}} & \multicolumn{2}{c}{\textbf{Quantum}} \\ \cline{3-2}  \cline{4-2}
         & & d=10 & d=13 \\
\hline
        50& 0001010000& 0001010000 & 0001010000   \\
       100 & 0011111000 & 0011111000 & 0011111000  \\
        200 & \textbf{0001010001} & 0001010010 & \textbf{0001010001}   \\
\bottomrule
    \end{tabular}
    \end{center}
    \label{T1}
\end{table}

By adding three additional layers, the outcomes from quantum computing agrees with the optimal solutions derived from classical methods. Hence, we deduce that the QNN circuit possesses superior expressive capabilities when $d=13$, which suffices for achieving optimal results. We employ 1000 shots for the optimization loop, allowing for the consistent output of a unique result. This implies that the same outcome is reproducible across 1000 measurements, confirming the result's precision and indicating that the solution accuracy is $100$\%.

\subsection{10-qubit exact solution}
We then examine a power system subject to dual constraints—namely, the power balance constraint and the spare capacity constraint—requiring the use of 5 auxiliary qubits. Due to the constraints of simulators, where the runtime for qubits exceeding 10 goes very large, we must reduce the number of units to 5, as the total number of qubits utilized is still 10.

\begin{figure}[H]
\centering
\includegraphics[width=0.6\textwidth]{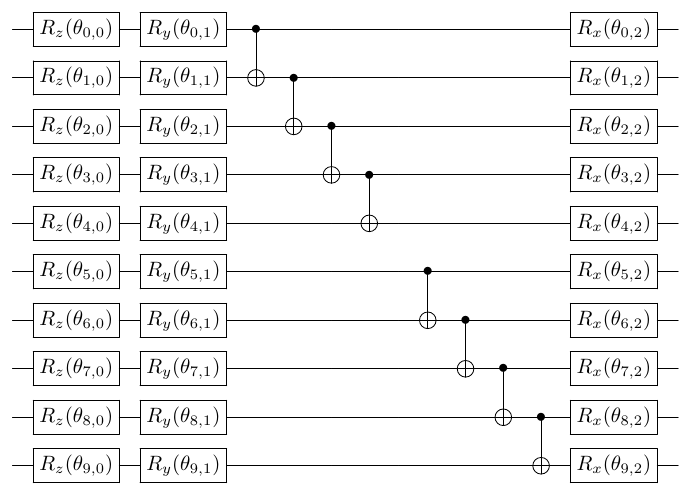}
\caption{Diagram of the knowledge-based PCQNN circuit for 5-unit system and double constraints (one layer).}
\label{PCQNNcircuit}
\end{figure}

In this 5-unit power system with dual constraints, auxiliary qubits are employed to represent the relaxation variables. To construct a PCQNN circuit, one can eliminate the CNOT gate connecting the 5th qubit and the 6th qubit, as discussed in section 3.2, where the auxiliary qubits are separated from the others. This configuration is depicted in Fig.\ref{PCQNNcircuit}

\begin{table}[H]
    \caption{Data for 5-unit power system.}
    \begin{center}
    \begin{tabular}{c|c|c|c|c|c}
    \toprule
     $i$ & $a_i(\$/MW^2)$ & $b_i(\$/MW)$ & $c_i(\$)$ & $p_i(MW)$ & $p_{i,max}(MW)$ \\
\hline
        1& 0.00175& 1.75 &0 &0.6&0.8  \\
       2 & 0.0625 & 1.00 & 0& 0.3& 0.5  \\
        3 &0.00834 & 3.25 & 0 &0.25& 0.35  \\
         4&0.0025 & 3.00 &0 &0.2& 0.3   \\
        5&0.0025 & 3.00 &0 &0.3& 0.4  \\
\bottomrule
    \end{tabular}
    \end{center}
    \label{5data}
\end{table}

\begin{table}[H]
    \caption{Comparison between quantum (including QNN and PCQNN) and classical algorithmw for 5-unit, double constraints}
    \begin{center}
    \scalebox{0.9}{
    \begin{tabular}{c|c|c|c|c|c|c|c|c}
    \toprule
        $L (MW)$ & $S (MW)$ & \textbf{HEA QNN} & probability & layers &\textbf{PCQNN} & probability & layers& \textbf{Classical}  \\
\hline
         0.6& 0.2 &01100 &0.83&7 & \textbf{10000} &1&6& \textbf{10000} \\
         0.9 & 0.2 & 11000& 1& 7& 11000 & 1&4&11000  \\
         1.1 & 0.4 & 11010 &1& 4& 11010 &1&4&11010  \\
         1.4 & 0.5 &11011 &1& 2 & 11011 &1&2&11011  \\
\bottomrule
    \end{tabular}}
    \end{center}
    \label{5-2}
\end{table}

Data for the 5-unit system can be located in Table~\ref{5data} \footnote{Generator data reference: K. Selvakumar et al., CSO Based Solution for Load Kickback Effect in Deregulated Power Systems, Applied Sciences.}. The input quantum state is initialized to the all-zero state, denoted as $\left | 0 \right \rangle$.
We proceed by employing the ansatz with both the hardware-efficient ansatz (HEA) QNN (as shown in Fig.\ref{fig2}) and the PCQNN (as depicted in Fig.\ref{PCQNNcircuit}) for comparative analysis. We also vary the number of layers, and the optimization outcomes are presented in Table~\ref{5-2}.
To comfirm the solutions, we obtain optimization results using a classical solver. The findings indicate that the solutions from the HEA QNN (based on the circuit in Fig.\ref{fig2}) closely match those of the classical solver, with the exception when $L=0.6$ and $S=0.2$. In this particular case, the QNN yields a solution of 01100 with a probability of 0.83. In contrast, the PCQNN achieves exact solutions across all cases with a probability of up to $100 \%$.
For the scenario with $L=0.9$ and $S=0.2$, the QNN requires at least 7 layers to achieve an exact solution, whereas the PCQNN only necessitates 4 layers. This demonstrates that the PCQNN can reduce the depth of the quantum circuit while maintaining precision, thereby enhancing computational efficiency.
Furthermore, as the scale of the units increases, we will inevitably encounter a greater number of inequality constraints. The proposed scheme offers a divisible quantum circuit, which opens up possibilities for distributed quantum computing.

\section{Conclusion}
We propose a QNN algorithm designed to address the UC problem. Our approach involves transforming the UC problem into a Quadratic Unconstrained Binary Optimization (QUBO) problem, which can be precisely resolved by the QNN quantum-classical hybrid algorithm.
In the case of a UC problem featuring 10-generators with a single constraint, the probability of achieving the optimal results is $100 \%$, thanks to the robust expressive capacity of a HEA QNN circuit with 13 repeating layers.
However, when dealing with a UC problem that includes 5-generators and two constraints, we observe that a PCQNN circuit outperforms the HEA QNN algorithm in terms of both accuracy and efficiency. By strategically eliminating entangled gates between unit qubits and auxiliary qubits based on the
symbolic domain knowledge, we have segmented the QNN circuit to create a divide-and-conquer quantum algorithm.
The PCQNN is a knowledge-based artificial quantum neural network, capable of constructing different elements through independent blocks. This quantum algorithm is not only suitable for the UC problem but also adaptable to a wide array of optimization problems characterized by diverse constraints.

\bigskip

\section*{Declaration of competing interest}
The authors declare that they possess no conflicting financial interests or personal relationships that could potentially bias the research findings presented in this paper.

\section*{Data availability statement}
The manuscript does not contain any accompanying data.

\section*{Acknowledgements}
\noindent This work is supported in part by the China Postdoctoral Science Foundation under Grant No. 2023M740874.

\bibliography{mybibfile}

\end{document}